\documentclass[conference,a4paper]{APSIPA2021}
\usepackage{multirow}
\usepackage{graphicx}
\usepackage{amsmath}
\usepackage[psamsfonts]{amssymb}
\usepackage{amsxtra}
\usepackage{threeparttable}
\usepackage{subfigure}
\usepackage{booktabs,cite}

\usepackage[backref]{hyperref}

\begin{document}

\title{Attention-based multi-channel speaker verification with ad-hoc microphone arrays}

\author{%
\authorblockN{%
Chengdong Liang, Junqi Chen, Shanzheng Guan and Xiao-Lei Zhang
}
\authorblockA{%
CIAIC, School of Marine Science and Technology, Northwestern Polytechnical University, Xi'an, China\\
E-mail: \{liangchengdong, jqchen, gshanzheng\}@mail.nwpu.edu.cn, xiaolei.zhang@nwpu.edu.cn}
%
}

\maketitle
\thispagestyle{empty}

\begin{abstract}
Recently, ad-hoc microphone array has been widely studied. Unlike traditional microphone array settings, the spatial arrangement and number of microphones of ad-hoc microphone arrays are not known in advance, which hinders the adaptation of traditional speaker verification technologies to ad-hoc microphone arrays. To overcome this weakness, in this paper, we propose attention-based multi-channel speaker verification with ad-hoc microphone arrays. Specifically, we add an inter-channel processing layer and a global fusion layer after the pooling layer of a single-channel speaker verification system. The inter-channel processing layer applies a so-called residual self-attention along the channel dimension for allocating weights to different microphones. The global fusion layer integrates all channels in a way that is independent to the number of the input channels. We further replace the softmax operator in the residual self-attention with sparsemax, which forces the channel weights of very noisy channels to zero. Experimental results with ad-hoc microphone arrays of over 30 channels demonstrate the effectiveness of the proposed methods. For example, the multi-channel speaker verification with sparsemax achieves an equal error rate (EER) of over $20\%$ lower than \textit{oracle one-best} system on semi-real data sets, and over $30\%$ lower on simulation data sets, in test scenarios with both matched and mismatched channel numbers.
\end{abstract}

\section{Introduction}
In the past decades, the performance of automatic speaker verification (ASV) has been improved significantly, such as the i-vector based methods \cite{dehak2010front} and deep neural network (DNN) based methods \cite{snyder2018x}.
However, such advances are mainly achieved on closed-talking scenarios with less interference. With the fast development of smart devices, such as smart speakers and various voice-enabled IoT gadgets, the need for far-field speech interaction will continue to grow. Recognizing who is speaking is essential to such smart devices for providing customized services. Far-field speech processing tasks including ASV remain challenging yet due to attenuated speech signals, noise interference, as well as room reverberations.

In order to solve the above challenging problem, various methods have been proposed at different stages of ASV systems. DNN based denoising methods \cite{wang2018supervised} for single-channel speech enhancement \cite{wang2013towards,pandey2019new} and multi-channel speech enhancement \cite{jiang2014binaural,heymann2016neural,wang2018all} were explored for ASV systems under complex environments \cite{taherian2020robust}. Linear prediction inverse modulation transfer function \cite{borgstrom2012linear} and weighted prediction error \cite{movsner2018dereverberation} methods were used for dereverberation. Warped minimum variance distortionless response (MVDR) cepstral coefficients \cite{jin2010speaker}, power-normalized cepstral coefficients (PNCC) \cite{kim2016power} and DNN bottleneck features \cite{yamada2013improvement} have been applied to ASV system for suppressing the adverse impacts of reverberation and noise.

The above research mainly focuses on single channel front-ends or multi-channel front-ends on single device. A microphone array with a known geometry is an important way to improve the performance of ASV. However, because speech quality degrades significantly when the distance between the speaker and microphone array enlarges, the performance of ASV is upper-bounded physically no matter how many microphones are added to the array \cite{chen2021scaling}. Compared with the fixed microphone array, an ad-hoc microphone array consists of a set of microphone nodes randomly placed in an acoustic environment \cite{zhang2021deep}. It provides more flexibility, and allows users to use their own mobile devices to virtually form a microphone array system. Recently, ad-hoc microphone arrays have been widely studied. \cite{yang2020deep} proposed deep ad-hoc beamforming based on speaker extraction for speech separation. In \cite{wang2021continuous}, a neural network architecture was proposed to capture both the inter-channel and temporal correlations from the multi-channel input of ad-hoc microphone arrays for speech separation. In \cite{qin2020hi}, the authors provided a set of baseline systems that are trained with the far-field speaker verification data in the transfer learning manner, and then performed the system fusion at the score level.

In this paper, we propose an attention-based multi-channel ASV with ad-hoc microphone arrays. Specifically, we first propose an inter-channel processing layer based on residual self-attention and a global fusion layer. Compared with single-channel ASV models, we add the inter-channel processing layer and the global fusion layer after the pooling layer. The softmax function in the residual self-attention of the  inter-channel processing layer transforms the similarity matrix to attention weights. Then, we further replace the softmax function with a so-called sparsemax function, which can force the channel weights of the noisy channels that do not contribute to the performance improvement to zero. We first train the single-channel ASV with clean speech data, then train the multi-channel ASV with ad-hoc data. This training strategy is motivated by the following two points: i) when we take some very noisy channels into training, the ASV system may not to be trained successfully; ii) the data of all channels is very large.

We conducted an extensive experiment on Librispeech simulated with ad-hoc microphone arrays and semi-real Libri-adhoc40 \cite{guan2021libri} corpora. Experimental result on the simulated dataset shows that the proposed multi-channel ASV with sparsemax achieves a relative EER reduction of $36.2\%$ over the \textit{oracle one-best} system, and $6.2\%$ over the multi-channel ASV with softmax on the matched 20-channel test scenario. Experimental result on the semi-real dataset shows that the proposed methods perform well in all test scenarios, and the model with sparsemax is slight better than the model with softmax.

\section{Proposed system}


In this section, we first introduce the single-channle ASV model. Then, we introduce the proposed attention-based multi-channle ASV. Finally, we introduce cross-channel multi-head  residual self-attention with sparsemax.

\subsection{Single-channel speaker verification system}
\begin{figure}[t]
	\begin{center}
		\includegraphics[width=0.3\linewidth]{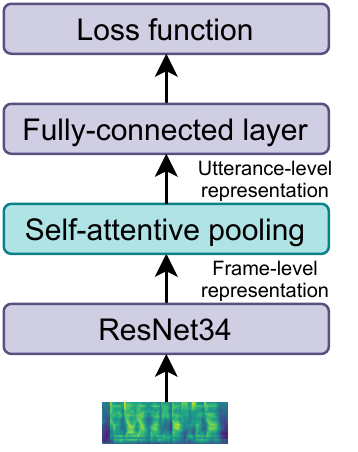}
	\end{center}
	\caption{Singel-channel speaker verification.}
	\label{fig:single_channel}
\end{figure}

As shown in Figure~\ref{fig:single_channel}, the network structure of the single-channel ASV is the same as in \cite{chung2020in}. The network architecture contains three main components: a front-end residual convolution neural network (ResNet) \cite{he2016deep}, a self-attentive pooling (SAP) \cite{zhu2018self} layer and a fully-connected layer. The front-end ResNet transforms the raw feature into a high-level abstract representation. The subsequent SAP layer outputs a single utterance-level representation. A fully-connected layer then further processes the utterance-level representation to be a more abstract utterance-level speaker embedding. All the components are jointly optimized in an end-to-end manner with a unified loss function. We adopt the angular prototypical loss \cite{chung2020in}. It constructs training batches in the same way as that in the original prototypical loss \cite{snell2017prototypical}.

\subsection{Multi-channel speaker verification system}
\begin{figure}
	\centering
	\includegraphics[width=0.4\linewidth]{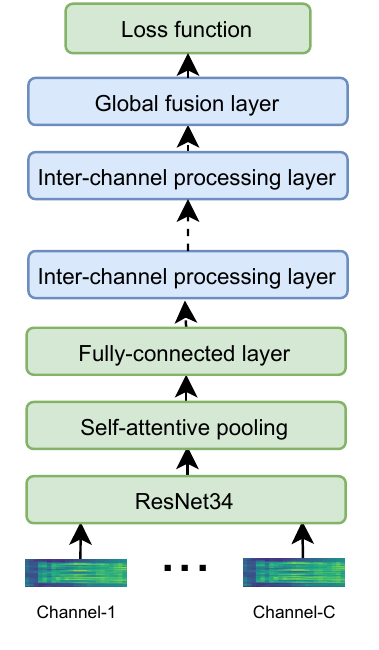}
	\caption{Multi-channel speaker verfication. The green block is first trained by a single-channel ASV on clean speech, and then fixed. The blue block is trained with the multichannel data collected from ad-hoc microphone arrays in noisy environments.}
	\label{fig:multi_channel}
\end{figure}

As shown in Figure~\ref{fig:multi_channel}, compared with the aforementioned single-channel system, the proposed multi-channel ASV system adds a novel inter-channel processing layer based residual self-attention and a global fusion layer after the SAP layer.

\begin{figure}[t]
	\centering
	\includegraphics[width=0.45\linewidth]{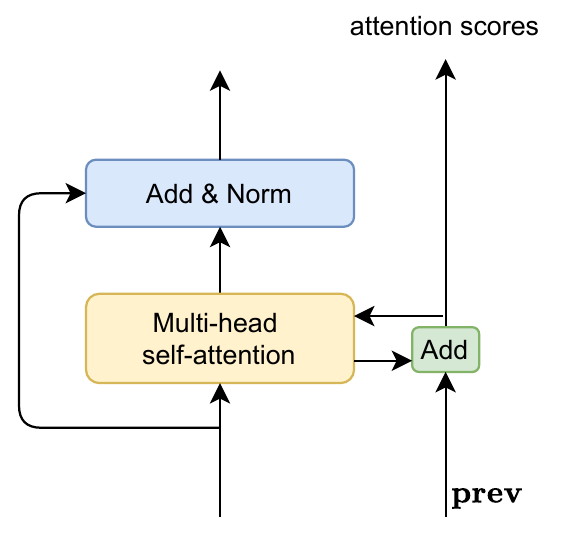}
	\caption{The residual self-attention layer.}
	\label{fig:resSA}
\end{figure}

\subsubsection{Inter-channel processing layer}
\begin{figure}
	\centering
	\subfigure[]{\label{fig:ch_attention}
		\includegraphics[width=0.35\linewidth]{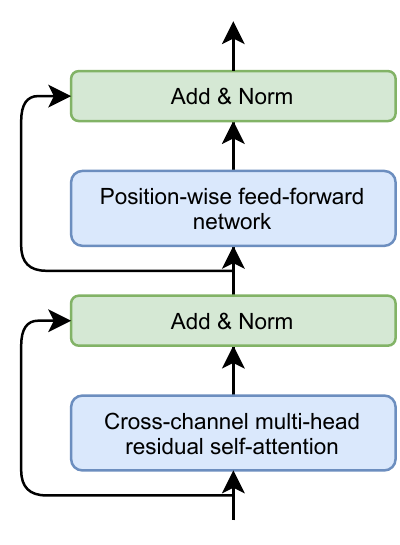}}
	\hspace{0.1\linewidth}
	\subfigure[]{\label{fig:global_fusion}
		\includegraphics[width=0.3\linewidth]{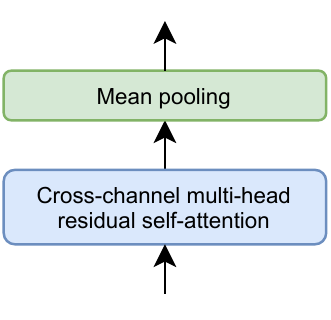}}
	\caption{(a) Inter-channel processing layer. (b) Global fusion layer. }
	\label{fig:subfig}
\end{figure}

The inter-channel processing layer collects the utterance-level representations from all channels, and outputs channel-reweighted utterance-level representations. As shown in Figure~\ref{fig:ch_attention}, the weights are calculated by a multi-head residual self-attention mechanism along the channel dimension, which is able to extract cross-channel information for the weight calculation. As shown in Figure~\ref{fig:resSA}, a residual self-attention layer takes the raw attention scores from previous layer as additive residual scores of the current attention function \cite{he2020realformer}. The detailed calculation process is as follows:

We stacked multiple inter-channel processing layers. For each layer, let $\mathbf{X}=\left[\boldsymbol{x}_{1}, \cdots, \boldsymbol{x}_{C}\right]$ denote the input, where $\boldsymbol{x}_{c} \in \mathbb{R}^{d}$ is an utterance-level feature of the $c$-th channel. while $C$ represents the number of channels and $d$ denotes the dimension of the feature. We assume that the number of the attention heads of the self-attention is $h$. For each attention head, the input features $\mathbf{X}$ are transformed into query, key, and value embedding subspaces of dimension $E$ respectively as follows:
\begin{equation}
\mathbf{Q}^{i}=\mathbf{X} \mathbf{W}_{Q}^{i}, \mathbf{K}^{i}=\mathbf{X} \mathbf{W}_{K}^{i}, \mathbf{V}^{i}=\mathbf{X} \mathbf{W}_{V}^{i}
\end{equation}
where $d_k = E /h$; the matrices $\mathbf{Q}$, $\mathbf{K}$, and $\mathbf{V}$ denote the query, key, and value embeddings respectively, all of which are in $\mathbb{R}^{C \times d_k}$; $\mathbf{W}_{k} \in \mathbb{R}^{d \times d_k}$ ($k \in \left\{K,Q,V\right\}$) are model parameters; and the superscript $i$ denotes the $i$-th attention head. Within each head, the cross-channel similarity matrix is obtained from the multiplication of the query and key matrices. A softmax function is applied to each column of the cross-channel similarity matrix to obtain an attention matrix $\mathbf{A}^{i} \in \mathbb{R}^{C \times C}$:
\begin{equation}
\mathbf{A}^{i}=\operatorname{softmax}\left(\frac{ \mathbf{Q}^{i} \cdot \left(\mathbf{K}^{i}\right)^\top}{\sqrt{d_k}} + \mathbf{prev} \right),
\label{eq:softmax_att}
\end{equation}
where $\mathbf{prev}$ is the attention scores from the previous inter-channel processing layer. Finally, the new attention scores $\frac{ \mathbf{Q}^{i} \cdot \left(\mathbf{K}^{i}\right)^\top}{\sqrt{d_k}} + \mathbf{prev}$ are sent to the upper layer.
The value matrix $\mathbf{V}^{i}$ is multiplied by the attention matrix as
\begin{equation}
\mathbf{H}^{i}=\mathbf{A}^{i} \cdot \mathbf{V}^{i},
\end{equation}
where $\mathbf{H}^{i} \in \mathbb{R}^{C \times d_k}$ is the output of  the $i$-th attention head, which is concatenated across the subspaces as:
\begin{equation}
\mathbf{Z} = \operatorname{Concat} \left[\mathbf{H}^{1}, \mathbf{H}^{2}, \ldots, \mathbf{H}^{h}\right] \mathbf{W}^{O}
\end{equation}
where $\mathbf{W}^{O} \in \mathbb{R}^{d \times d}$ is the weight matrix of the linear projection layer.

The position-wise feed-forward network (FFN) with ReLU activation is applied to $\mathbf{Z}$ to generate the output of the current inter-channel processing layer. A residual connection is applied between the input and output of the cross-channel multi-head self-attention layer as well as the FFN layer to mitigate the gradient vanishing problem.

\subsubsection{Global fusion layer}
After multiple layers of inter-channel processing, the global fusion layer is added above the top inter-channel processing layer. As shown in Figure~\ref{fig:global_fusion}, it is implemented by a cross-channel multi-head residual self-attention module, followed by a mean pooling operator. It fuses the information of all channels in a way that is independent to the number of the input channels.

\subsection{Cross-channel multi-head self-attention with sparsemax}
The common softmax has a limitation for ad-hoc microphone arrays that the elements of its output can never be zero, which cannot be used for channel selection. As we know, if some channels are very noisy, they may hurt the system when their weights are nonzero. To address this problem, we replace softmax with sparsemax \cite{martins2016softmax} in \eqref{eq:softmax_att}, where sparsemax is defined as:

\begin{equation}
\operatorname{Sparsemax}(\boldsymbol{z})=\underset{\boldsymbol{p} \in \Delta^{K-1}}{\arg \min }\|\boldsymbol{p}-\boldsymbol{z}\|^{2}
\end{equation}
where $\Delta^{K-1}=\left\{\boldsymbol{p} \in \mathbb{R}^{K} \mid \sum_{i=1}^{K} p_{i}=1, p_{i} \geq 0\right\}$ represents a $(K-1)$-dimensional simplex. Sparsemax will return the Euclidean projection of the input vector $\boldsymbol{z}$ onto the simplex, which is a sparse vector. Its solution has the following closed-form:

\begin{equation}
\operatorname{Sparsemax}_{i}(\boldsymbol{z})=\max \left(z_{i}-\tau(\boldsymbol{z}), 0\right)
\end{equation}
where $\tau$ : $\mathbb{R}^K \rightarrow \mathbb{R}$ is a function to find a soft threshold. Let $ z_{(1)} \geq z_{(2)} \geq \ldots \geq z_{(K)}$ be the sorted coordinates of $\boldsymbol{z}$, and define $k(\boldsymbol{z}):= \max \left\{k \in[K] \mid 1+k z_{(k)}>\sum_{j \leq k} z_{(j)}\right\}$. Then,
\begin{equation}
\tau(\boldsymbol{z})=\frac{\left(\sum_{j \leq k(\boldsymbol{z})} z_{(j)}\right)-1}{k(\boldsymbol{z})}.
\end{equation}

\section{Experiments}
\subsection{Dataset}

Our experiments use three data sets, which are the Librispeech corpus \cite{panayotov2015librispeech}, Librispeech simulated with ad-hoc microphone arrays (Libri-adhoc-simu), and Librispeech played back in real-world scenarios with 40 distributed microphone receivers (Libri-adhoc40) \cite{guan2021libri}. Librispeech contains more than 1000 hours of read English speech. In our experiments, we selected 960 hours of data including 2338 speakers to train single-channel ASV system, and selected 10 hours of data including 40 speakers for development.

Libri-adhoc-simu uses 'train-clean-100' subset of the Librispeech data as the training data, which contain 251 speakers. It uses 'dev-clean' subset as development data, and takes 'test-clean' subset as test data, which contain different 40 speakers respectively. For each utterance, we simulate a room. The length and width of the room are selected randomly from a range of $\left[5,25\right]$ meters. The height is selected randomly from $\left[2.7, 4\right]$ meters. Multiple microphones and one speaker source are placed randomly in the room. We constrain the distance between the source and the walls to be greater than $0.2$ meters, and the distance between the source and the microphones to be at least $0.3$ meters \cite{chen2021scaling}. We use an image-source model\footnote{https://github.com/ehabets/RIR-Generator} to simulate a reverberant environment and selected $T_{60}$ from a range of $\left[0.2, 0.4\right]$ second. A diffuse noise generator\footnote{https://github.com/ehabets/ANF-Generator} is used to simulated uncorrelated diffuse noise. The noise source for training and development is a large-scale noise library containing over $20000$ noise segments \cite{tan2021speech}, and the noise source for test is the noise segments from CHiME-3 dataset \cite{barker2015third} and NOISEX-92 corpus \cite{varga1993assessment}. We randomly generate $20$ channels for training and development, and $20$, $30$ and $40$ channels respectively for test.

Libri-adhoc40 is collected by playing back the 'train-clean-100', 'dev-clean', and 'test-clean' subset of Librispeech in a large room \cite{guan2021libri}. The recording environment is a real office room with one loudspeaker and 40 microphones. It has strong reverberation with little additive noise. The distances between the loudspeaker and microphones are in a range of $\left[0.8,7.4\right]$ meters. The positions of the loudspeaker and microphones are different in the training and test set. We randomly select $20$ channels for each training and development utterances, and $20$, $30$, $40$ channels for each test utterance which corresponds to three test scenarios.

\subsection{Model structure}
Figure~\ref{fig:single_channel} shows the architecture of the proposed single-channel model. We set the widths of the residual blocks to $\{16,32,64,128\}$. The embedding size in the fully-connected layer is $512$. As shown in Figure~\ref{fig:multi_channel}, multi-channel speaker verification system adds the Inter-channel processing layer and the Global fusion layer after the SAP layer. Four inter-channel processing layers are stacked. The output dimensions of each self-attention layer and FFT layer are both 256. The number of the attention heads in each attention layer is 4. We use voxceleb\_trainer\footnote{https://github.com/clovaai/voxceleb\_trainer} to build our models.

\subsection{Model training}
During training, we use a fixed length 2 second temporal segment, extracted randomly from each utterance. A Hamming window with a width of 25ms and a step length of 10ms is used to extract the spectrum. The 40-dimensional Mel filterbanks are used as the input. Mean and variance normalisation (MVN) is performed by applying instancde normalisation \cite{ulyanov2016instance} to the network input.

For each epoch, we randomly sample a maximum of $100$ utterances from each speaker to reduce class imbalance. No data augmentation is performed during training, apart from the random sampling. We use the Adam optimizer with an initial learning rate of $0.001$ decreasing by $5\%$ every $10$ epochs. First, the single-channel ASV system is trained for $200$ epochs on the Librispeech corpus. Then, the parameters of the single-channel ASV are fixed and sent to the multi-channel ASV. Finally, we trained the multi-channel ASV system with Libri-adhoc-simu data and Libri-adhoc40 data respectively.

After training, we sample five 4-second temporal crops at regular intervals from each test segment, and compute the similarities between all possible combinations ($5 \times 5 = 25$) from each pair of segments. The mean of the $25$ similarities is used as the score. We arrange all utterances in test set to get $6861780$ trails for testing, including $183922$ positive trails and $6677858$ negative trails.

\subsection{Results}
We compare the proposed multi-channel ASV with softmax and sparsemax. Moreover, we construct an \textit{oracle one-best} baseline, which picks the channel that is physically closest to the sound source as the input of the single-channel ASV model. Note that, for the \textit{oracle one-best} baseline, the distances between the speaker and microphones are known beforehand.

Table~\ref{tab:simu} lists the preformance of the comparison methods on Libri-adhoc-simu. From the table, we see that all of the proposed methods perform well in both test scenarios. Particularly, the multi-channel ASV with softmax achieves an EER of $31.9\%$ lower than \textit{oracle one-best} baseline on the 20-channel test scenario, and $27.7\%$ on the mismatched 30-channel test scenario. The generalization performance in the mismatched 30-channel and 40-channel test environment is even better than the performance in the matched 20-channel environment, which demonstrates the advantage of adding channels to ad-hoc microphone arrays. Sparsemax achieves significant performance improvement over softmax. For example, sparsemax achieves a relative EER reduction of $6.2\%$ over softmax on the matched 20-channel.

Table~\ref{tab:real} lists the results on the Libri-adhoc40 semi-real data. From the table, we see that the proposed model performs well. The model with softmax achieves a relative EER reduction of $35.3\%$ over the \textit{oracle one-best} baseline on the 20-channel test scenario, and $26.3\%$ on the mismatched $30$-channel test scenario. Sparsemax is slightly better than softmax, with a relative EER reduction of $3\%$ in all test scenarios. The above results show the proposed models is effective for ad-hoc microphone arrays.

\begin{table}[t]
	\begin{center}
		\begin{threeparttable}
			\caption{EER($\%$) comparison on \textbf{Libri-adhoc-simu}. The term "ch" is short for channels in test. MC-ASV is short for multi-channel ASV.}
			\begin{tabular}{llll}
				\toprule
				Method	& 20-ch 	& 30-ch	& 40-ch \\
				\midrule
				Oracle one-best 	& 11.90		& 10.99		& 10.77		\\
				MC-ASV with softmax (proposed) & 8.10		& 7.94		& 7.88 \\
				MC-ASV with sparsemax (proposed) & 7.59 & 7.53 & 7.47 \\
				\bottomrule
			\end{tabular}
			\label{tab:simu}
		\end{threeparttable}
	\end{center}
\end{table}

\begin{table}[t]
	\begin{center}
		\begin{threeparttable}
			\caption{EER($\%$) comparison on \textbf{Libri-adhoc40} semi-real data.}
			\begin{tabular}{llll}
				\toprule
				Method	& 20-ch 	& 30-ch	& 40-ch \\
				\midrule
				Oracle one-best 	& 17.29			& 14.86		& 13.56		\\
				MC-ASV with softmax (proposed) & 11.18	& 10.95		& 10.84			\\
				MC-ASV with sparsemax (proposed) & 10.89 & 10.62 & 10.53 \\
				\bottomrule
			\end{tabular}
			\label{tab:real}
		\end{threeparttable}
	\end{center}
\end{table}

\section{Conclusions}
In this paper, we first present the multi-channel ASV model with ad-hoc microphone arrays using the inter-channel processing layer based residual self-attention and global fusion layer, which is proposed to take advantage of multi-channel information in a way that is independent of the number and permutation of the microphones. The inter-channel processing layer, which aims to learn the channel weights, consists of a residual self-attention and FFN. Multiple inter-channel processing layers are stacked, followed by a mean pooling layer for global information fusion. Then, we future replace the softmax opterator in the residual self-attention with sparsemax, which forces the channel weights of the noisy channels to zero. We evaluate our model on Libri-adhoc-simu with background noise and Libri-adhoc40 with high reverberation. Experimental results show that the proposed multi-channel ASV with sparsemax outperforms that with softmax and the oracle baseline in both simulated data and semi-real corpus. The results also demonstrate the importance of channel selection to ASV with large-scale ad-hoc microphone arrays.

\bibliographystyle{IEEEtran}

\bibliography{reference}

\end{document}